\documentclass[Afour,sageh,times]{sagej}
\usepackage[colorlinks,bookmarksopen,bookmarksnumbered,citecolor=red,urlcolor=red]{hyperref}
\usepackage{moreverb,url}

\newcommand\BibTeX{{\rmfamily B\kern-.05em \textsc{i\kern-.025em b}\kern-.08em
T\kern-.1667em\lower.7ex\hbox{E}\kern-.125emX}}
\usepackage{cprotect}
\usepackage{adjustbox}
\usepackage{flushend}
\usepackage{float}
\usepackage[usenames,dvipsnames]{xcolor}
\usepackage{braket}
\usepackage{times}
\usepackage{float}
\usepackage{stfloats}   
\usepackage{upquote,textcomp}
\usepackage{multicol}
\usepackage{algorithm2e}
\usepackage{multirow}
\def\volumeyear{202X}
\def\journalname{~}

\usepackage{url}
\usepackage{hyperref}
\usepackage[hyphenbreaks]{breakurl}

\begin{document}
\runninghead{Belfore}
\pagestyle{plain}

\title{A Scalable FPGA Architecture for Quantum Computing Simulation}
\author{Lee A. Belfore II}
\affiliation{Old Dominion University, USA}
\corrauth{Lee A. Belfore II, Old Dominion University,
  Norfolk, VA 23323, USA.}

\email{lbelfore@odu.edu}

  \begin{abstract}
    A quantum computing simulation provides the opportunity to explore
    the behaviors of quantum circuits, study the properties of quantum
    gates, and develop quantum computing algorithms.  Simulating
    quantum circuits requires geometric time and space complexities,
    impacting the size of the quantum circuit that can be simulated as
    well as the respective time required to simulate a particular
    circuit. Applying the parallelism inherent in the simulation and
    crafting custom architectures, larger quantum circuits can be
    simulated.  A scalable accelerator architecture is proposed to
    provide a high performance, highly parallel, accelerator.  Among
    the challenges of creating a scalable architecture is managing
    parallelism, efficiently routing quantum state components for gate
    evaluation, and measurement. An example is demonstrated on an
    Intel Agilex field programmable gate array (FPGA).
\end{abstract}
\makeatletter
\thispagestyle{empty}
~\\
\begin{figure*}[ht]
\begin{sf}
  {\raggedright\titlesize\textbf{\@title} \par}%
  ~\\[3em]

{\par\large%
{\raggedright\textbf{\@author}\par}}
{\raggedright{Old Dominion University, Norfolk, VA USA}\par}
{\raggedright{lbelfore@odu.edu}\par}
\vskip 40pt%
{\noindent\usebox\absbox\par}
{\vspace{20pt}%
{\noindent\normalsize\@keywords}\par}
\end{sf}
\makeatother
\end{figure*}
\vspace*{-40pt}
\section{Introduction}
Quantum computing continues to grow in interest because of its
theoretical ability to solve intractably hard computational problems
that are beyond the capabilities of conventional computing methods
\cite{Shor-1994,Grover-1996}. Concurrently, research and industry
efforts have succeeded in demonstrating increasingly capable quantum
computing platforms \cite{Dwave-2020,IBM-Q-2019,Arute-2019}. Of note,
quantum supremacy \cite{preskill12} is the point where the quantum
computing platform is able to solve problems that classical systems
are not capable. Presently, the reported largest quantum computer
implementations vary according to the platform \& inherent
capabilities and continues to increase as technological hurdles are
crossed (for example \cite{IBM-53-qubit,Arute-2019,cao-23}) and are at
the threshold of quantum supremacy.

Because access to quantum computing platforms have limited or
restricted access, quantum computing simulation provides a means for
people to learn about quantum computing and test their ideas within
the limits of the simulator's capabilities. Further, algorithms and
other ideas that may require features not present in existing quantum
computing platforms, for example a novel quantum gate that is an
important algorithmic building block.

Many capable quantum computing simulators are available
\cite{Gheorghiu-2018,QuEST-2019,xu23}.  QuEST \cite{QuEST-2019}, like
other simulation platforms, includes the ability to parallelize
simulations, utilizing CPU cores and GPUs, for increasingly larger
problem sizes. QuEST was demonstrated to be capable of simulating
quantum circuits with 38 qubits on a 2,048 node supercomputer.

Theoretically, in order to simulate any general quantum circuit, the
simulator must include the ability to include gates that form a
universal set \cite{Gottesman98,Boykin99}.  As noted in
\cite{Gottesman98}, gates within the Clifford group are not fully
universal and must be augmented with additional gates to make the set
universal. Of importance in this work, \cite{Boykin99} describes a set
of universal gates consisting of one and two input gates.

The approach described here presents a demonstration of a quantum
computing simulator implemented on a high performance field
programmable gate array (FPGA) platform. FPGAs offer flexible
programming available in a large and capable programmable fabric that
offers the equivalent of several million logic functions. In addition,
digital signal processing (DSP) units provide optimized computational
modules in capacities of thousands and more. Each DSP is potentially
capable of completing two multiplications every clock cycle at clock
frequencies exceeding 600MHz \cite{Agilex-video}.  Further, block
random access memory (BRAM) provides fast, flexible memory that can
feed DSPs with data at high rates. Finally, FPGAs, particularly high
end devices, offer high speed networking and PCIe interfaces to
support distributed applications.

Building on these raw resources, FPGAs provide a flexible platform to
construct arbitrary computational structures that can take leverage
the unique features of a computational problem that conventional CPU
\& GPU based platforms cannot. The parallelism in the problem can be
more directly matched to the computational problem as well as the flow
of data and computational products.  For example, bandwidth issues
related to memory hierarchies can be minimized in FPGA implementations
by placing memory containing data operands close to the computational
functional units.

Several approaches for simulating quantum computers using FPGAs are
described in the literature.  Early results are limited as a result of
the technology of the time.  An early result, Khalid et
al. \cite{Khalid-2004} describes an approach for specifying and
synthesizing quantum circuit simulators. Khalid's approach is flexible
in that an arbitrary circuit can be specified, within the capabilities
of the target FPGA.  A weakness in Khalid's approach is that
simulating a new circuit will require specifying the desired circuit
and then synthesizing which can be time consuming, particularly for
larger circuits.  Frank et al. \cite{Frank09} describes an approach
for implementing an FPGA based quantum computing simulator but did not
provide an actual demonstration implementation.  Similar to Khalid,
Silva and Zabeleta \cite{silva17} similarly synthesizes a specific
circuit, the quantum Fourier transform (QFT) which is an important
building block in many important quantum algorithms. Of note, the
authors use Xilinx High Level Synthesis (HLS) \cite{Xilinx-HLS} to
specify the circuit that is subsequently synthesized. As with
\cite{Khalid-2004}, each new circuit must be synthesized which, for
large circuits, may incur a significant amount of time for synthesis.
Pilch and D{\l}ugopolski \cite{Pilch-2019} propose a general FPGA
based emulator that includes the marrying of quantum and classical
procedural computing. Furthermore, their approach supports general
two-input quantum gates and includes a demonstration of the Deutsch
gate \cite{Deutsch89}. While the analysis provided in the paper is
comprehensive, the demonstration was for a two-qubit circuit example.
\cite{Suzuki23} presents the application of a special purpose hybrid
CPU-FPGA quantum computing simulator that simulates quantum machine
learning using the quantum kernel method for image
classification. Notably, the simulation demonstrated the use of a
6-bit quantum kernel applied to a quantum support vector machine
(QSVM).

In this paper, we present a scalable FPGA based quantum computing
simulator architecture. The architecture supports any one-input
quantum gate, and two-input quantum gates where the magnitude of
matrix elements is one. The gates supported are consistent with
universal quantum computing. The simulator is able to take an
arbitrary circuit specification within the noted gate restrictions.
The simulator has been implemented on an Intel Agilex FPGA capable of
modeling arbitrary seven-bit quantum circuits. The results are
functionally validated by comparing against the QuEST quantum logic
simulator. Attempt to directly leverage FPGA resources, parallelize

This paper is organized into seven sections including
an introduction,
an overview of quantum circuit simulation,
a review of important architectural decisions,
a detailed discussion of the quantum simulation unit (QSU),
a presentation of examples, and a summary \& future work.
\section{Simulating Quantum Circuits}
In the context of this paper, quantum circuits are simulated on
classical computing platforms.  In this section, aspects of quantum
circuit simulation important to this paper are present and an in depth
treatment of quantum computing simulations can be found elsewhere
\cite{QuEST-2019,Qiskit}. The relevant parameters, operations, and 
essential properties are described in this section.

\subsection{Quantum State}
The representation of the quantum state is central to the simulation
of quantum circuits. The state of an $n$-bit quantum circuit is
represented by a vector of $2^n$ probability amplitudes, where the
$i^{\mbox{th}}$ component holds the probability amplitude for that
state. This quantum state vector (QSV) is capable of representing an
arbitrary entangled state.
Apropos, the QSV for an $n$-qubit circuit gives a
snapshot of the state of the quantum state at some point in time. The
QSV is expressed as
\begin{equation}
\ket{\Psi}=\sum_{i=0}^{2^{n-1}} \alpha_i \ket{b_i}
\label{eq:n-qubit-state-1}
\end{equation}
where $n$ is the number of qubits, $b_i$ is the $n$-bit binary code
for state $i$, $\ket{b_i}$ is the state, and $\alpha_i$ is the
i$^{\mbox{\scriptsize{th}}}$ complex probability amplitude. Note the
geometric relationship between $n$ and the number of components in
the quantum state.
Further, the norm of the state vector must be unity, or
\begin{equation}
  \left|
  \rule[-0.38\baselineskip]{0pt}{0.38\baselineskip}
       {\ket{\Psi}}\right|=\sum_{s=0}^{2^n-1} |{\alpha_i}|^2=1.
\end{equation}

\subsection{Simple Introduction to Quantum Circuit Operation}
\label{ssc:quantum-computing}
Here, a simple introduction to quantum circuit operation on one qubit
is presented. The interested reader can consult
\cite{Nielsen10} for further details.
In
classical computing, information is represented as bits taking on the
values \verb_'0'_ and \verb_'1'_. Quantum information, on the other
hand, is represented by qubits that are entangled mixtures of
\verb_'0'_s and \verb_'1'_s. Individual qubits are represented by a
one-dimensional array with two elements giving the entangled
contributions for each of the binary values. The ``ket'' notation is
used to describe the states.
\begin{equation}
\ket{0} = \left[ \begin{array}{r} 1 \\ 0\end{array}  \right] \qquad
\ket{1} = \left[ \begin{array}{r} 0 \\ 1\end{array}  \right].
\label{eq:sharp-binary}
\end{equation}
Equation (\ref{eq:sharp-binary}) describes what are termed ``sharp'' values. These
are values that are known with certainty and contain no entanglements
with other values or qubits.

More generally, a qubit can be described by the following equation
\begin{equation}
\ket{\Psi}=\alpha \ket{0} + \beta \ket{1},
\label{eq:one-qubit-state}
\end{equation}
where $\alpha$ and $\beta$ are complex values such that
$|\alpha|^2+|\beta|^2=1$.
Equation (\ref{eq:one-qubit-state}) represents a value is neither \verb_'0'_ or
\verb_'1'_ and that it is a quantum entanglement of these values.
The entanglement can be extended to multiple bits. The following state
gives the general form for the entangled state of two qubits
\begin{equation}
  \ket{\Psi}=\alpha \ket{00} + \beta \ket{01} + \gamma \ket{10} + \delta \ket{11},
\label{eq:two-qubit-state}
\end{equation}
where similarly $|\alpha|^2+|\beta|^2+|\gamma|^2+|\delta|^2=1$.

A general one qubit gate operation can be defined as an arbitrary
rotation on the Bloch Sphere as the unitary
matrix that follows from
\begin{equation}
  G_1=U=e^{j\alpha}R_z(\beta)R_y(\gamma)R_z(\delta).
\label{eq:one-qubit-gate-unitary}
\end{equation}
In the context of the Bloch sphere, 
$\alpha$ is the rotation about the $x$-axis, 
$R_y(\cdot)$ is a rotation about the $y$-axis, 
$R_z(\cdot)$ is a rotation about the $z$-axis, and $\beta$, $\gamma$, \& $\delta$ and the
respective rotation angles
\cite{Nielsen10}. The unitary matrix $G_1$ will designate a one-input
quantum gate. 
In the general case, the matrix elements are complex.
A one-qubit gate operation on a qubit is described as
\begin{equation}
  \ket{\Psi'}=U \ket{\Psi}
  \label{eq:one-qubit-gate-operation}
\end{equation}
where $\ket{\Psi'}$ is the updated state after performing the gate
operation. Two input gates are similarly formulated.

\subsection{Universal Set of Gates}

The gates available in the simulator must be carefully considered so
as to provide a universal set of gates.
The Clifford group, consisting of the Hadamard Gate ($\mathit{H}$), the
S~gate ($\mathit{S}$), and and the controlled-not gate ($\mathit{CNOT}$) gate,
form the basis for many useful results but is not universal
\cite{Gottesman98}. In order to meet the criteria in the previous
paragraph, the most suitable universal set is meeting the criteria of
one and two input gates is given by \cite{Boykin99}, which consists
of the gates $\mathit{H}$, $\mathit{CNOT}$, and the T~gate ($\mathit{S}$). 
Beyond the minimum set of gates required for universal quantum
computation, additional gates provide a means for representing quantum
circuits in a more compact fashion.

\subsection{Evaluation of a Quantum Gate on a Classical Computer}

On a classical computer, the gate applied to the $2^n$ length quantum
state $\ket{\Psi}$ requires ``touching'' each element in the
vector. Assuming a one-input quantum gate applied to a qubit $q_i$,
and further assuming the qubits are permuted so that $q_i$  is now
the right-most, $\ket{\Psi^i}$. Successive pairs of elements in
$\ket{\Psi^i}$
are identical in the first $(n-1)$ qubits state and
differ in the last, nominally representing the state vector subset
{\small $\left[ \begin{array}{c}  0 \\ 1 \end{array} \right]$}.
Applying the gate matrix to this pair
of states gives the updated state given the first $(n-1)$ are
constant. Repeating for all $2^{n-1}$ pairs from the quantum state
register results in an updated state vector reflecting the application
of the gate to the original state vector. Two~input gates can be
similarly formulated, with a permuted state $\ket{\Psi_n^{i,j}}$, with
the two-input quantum gate applied to successive quartets from the
permuted quantum state register. The original quantum state is
restored by applying a permutation consistent with the original
ordering of the qubits. Notably, provided an efficient permutation
operation and sufficient parallelism is available, performing gate
operations is straightforward, with the caveat that the required
parallelism is $O(2^n)$.
Pseudocode for evaluating gates is outlined in
Algorithms~\ref{alg:one-input-gate}~and~\ref{alg:two-input-gate}.

\begin{table}
  \begin{center}
  \begin{tabular}{|@{\hspace{0.25em}}l@{\hspace{0.25em}}|@{\hspace{0.25em}}p{2.4in}@{\hspace{0.25em}}|}
    \hline
    \bf Symbol & \bf Definition \\ \hline \hline
    $\ket \Psi$ & QSV \\ \hline
    $\ket \Psi_i$ & $i^{\mbox{\scriptsize th}}$ component of QSV \\ \hline
    $\ket \Psi_{i..j}$ & QSV segment from $i$ to $j$ \\ \hline
    $\ket {\Psi^i}$ & permuted QSV, qubit $i$ is LSB \\ \hline
    $\ket {\Psi^{ij}}$ & permuted QSV, qubit $i$ is LSB, and
    $j$ next most \\ \hline
    $\psi$      & unnormalized segment of a QSV \\ \hline
    $\pi(\ket \Psi,i)$     & quantum state permutation resulting from
    moving qubit $i$ to the LSB. Result of permutation is$\ket {\Psi^i}$  \\ \hline
    $\pi^{-1}(\ket {\Psi^i},i)$     &  inverse permutation restoring the
    state vector to its original ordering\\ \hline
    $G_i$      & $2^i \times 2^i$ unitary matrix defining function for
    an $i$ input gate \\ \hline
    $\{G_1,i\}$ & Specification of one input gate applied to qubit $i$\\ \hline
    $\{G_2,i,j\}$ & Specification of two input gate applied to qubits $i$, $j$\\ \hline
     ${\bf C}$      & Quantum circuit consisting of a Sequential list of one \& two input gate
    specifications, $\{G_1,i\}$ or $\{G_2,i,j\}$, that defines a
    quantum circuit
   quantum gate  \\ \hline
  \end{tabular}
  \end{center}
  \caption{Notation used in algorithmic specifications}
\end{table}
\begin{algorithm}
    \SetKwInOut{Inputs}{Inputs}
    \SetKwInOut{Output}{Output}

    \underline{QuantumOne} $(\ket{\Psi}, G_1, i )$\\
    \Inputs{Initial quantum state $\ket{\Psi}$\\
      One-input gate $G_1$\\
      Qubit $i$
    }
    \Output{Updated quantum state $\ket{\Psi'}$}

    // does not include measurement gate\\
    $\ket{\Psi^i}$=$\pi$($\ket{\Psi},i$)\\
    \For {$ c \in \{0,1,...,2^{n-1}\}$}
    {
      $\psi_c$=${\ket{\Psi^{i}}}_{2c\, .. \, 2c+1}$\\
      $\psi_c ' $=$G_1\psi_c$\\
      ${\ket{\Psi^{i}}'}_{2c\, .. \, 2c+1}$=${\psi_c '}$\\
      }
    $\ket{\Psi '}$=$\pi^{-1}$(${\ket{\Psi^i}'},i$)\\
    return $\ket{\Psi '}$\\
    \vspace*{1em}
    \caption{One-input quantum gate evaluation}
    \label{alg:one-input-gate}
\end{algorithm}

\begin{algorithm}
    \SetKwInOut{Inputs}{Inputs}
    \SetKwInOut{Output}{Output}

    \underline{QuantumTwo} $(\ket{\Psi}, G_2, i,j )$\\
    \Inputs{Initial quantum state $\ket{\Psi}$\\
      Two-input gate $G_2$\\
      Qubits $i$, $j$
    }
    \Output{Updated quantum state $\ket{\Psi'}$}

    $\ket{\Psi^{ij}}$=$\pi$($\ket{\Psi},i,j$)\\
    \For {$ c \in \{0,1,...,2^{n-2}\}$}
    {
      $\psi_c$=${\ket{\Psi^{ij}}}_{4c\, .. \, 4c+3}$\\
      $\psi_c ' $=$G_2\psi_c$\\
      ${\ket{\Psi^{ij}}'}_{4c\, .. \, 4c+3}$=${\psi_c '}$\\
      }
    $\ket{\Psi '}$=$\pi^{-1}$($\ket{{\Psi^{ij}}'},i,j$)\\
    return $\ket{\Psi '}$\\
    \vspace*{1em}
    \caption{Two-input quantum gate evaluation}
    \label{alg:two-input-gate}
\end{algorithm}

A measurement gate, or M~gate, is fundamentally different in that it
is not reversible and results in a discrete value, either 0 or 1, for
the measured qubit. The M~gate also is the means for retrieving
results from the quantum circuit. Measuring a qubit on a classical
computer requires three steps: quantum state.  Note that the
measurement gate requires examining all elements of the QSV 
twice. Further, summing the probabilities serializes the
function of the gate, limiting the beneficial aspects of providing
hardware parallelism to complete the operation.  The algorithm for
evaluating an M-gate is given in Algorithm~\ref{alg:measurement-gate}.

\begin{algorithm}
    \SetKwInOut{Inputs}{Inputs}
    \SetKwInOut{Output}{Output}

    \underline{Measure} $(\ket{\Psi}, i )$\\
    \Inputs{Initial quantum state $\ket{\Psi}$\\
      Qubit measured $i$
    }
    \Output{Updated quantum state $\ket{\Psi '}$}

    $\ket{\Psi^i}$=$\pi$($\ket{\Psi},i$)\;
    $P_0$=0;                // Initialize zero probability\\
    \For {$ c \in \{0,1,...,2^{n-1}\}$}
    {
      $P_0$=$P_0+|\ket{\Psi^i}_{2c}|^2$
    }

    \eIf{$P_0\ne 0$ {\rm and} $P_0 \ne 1$}
       {
         prn=Rand(0,1);  // uniform random number in [0,1]\\
         \For{$ c \in \{0,1,...,2^{n-1}\}$} {
           \eIf{{\rm prn} $<P_0$}
               {\noindent
                 // Qubit measured as zero\\
                 $\ket{{\Psi^i}'}_{2c}$=${\ket{\Psi^i}}_{2c}/\sqrt{P_0}$\\
                 $\ket{{\Psi^i}'}_{2c+1}$=$0$
               }
               {
                 // Qubit measured as one\\
                 $P_1$=$1-P_0$\\
                 $\ket{{\Psi^i}'}_{2c}$=$0$\\
                 $\ket{{\Psi^i}'}_{2c+1}$=${\ket{\Psi^i}}_{2c+1}/\sqrt{P_1}$
               }
         }
       }
       {
         ${\ket{\Psi^i}'}$=$\ket{\Psi^i}$\\
       }
    $\ket{\Psi '}$=$\pi^{-1}$($\ket{{\Psi^i}'},i$)\\
    return $\ket{\Psi '}$\\
    \vspace*{1em}
    \caption{Measurement gate}
    \label{alg:measurement-gate}
\end{algorithm}

A quantum circuit is evaluated by sequentially applying quantum logic
gates to the QSV.  a sequence of gate operations applied to
qubits. In the context of simulation, this naturally means that the
gate operations are applied to the QSV as described
in the previous section. Algorithm~\ref{alg:quantum-circuit} gives the
process for evaluating a quantum circuit using the previously defined algorithms.
Taken together,
Algorithms~\ref{alg:one-input-gate}-\ref{alg:quantum-circuit} define
the process for simulating a quantum circuit on a classical computer.

\begin{algorithm}
    \SetKwInOut{Inputs}{Inputs}
    \SetKwInOut{Output}{Output}

    \underline{EvaluateCircuit} $(\ket{\Psi}, i )$\\
    \Inputs{Initial quantum state $\ket{\Psi}$\\
      Quantum circuit $\bf C$
    }
    \Output{Updated quantum state $\ket{\Psi '}$}
    // assume list of quantum gates is a vector of length $C$\\
    $\ket{\Psi '}$=$\ket{\Psi}$\\
    \For {$ g \in \{0,1,...,C-1\}$}
    {
      \eIf{${\bf C}[g]${\rm .gate} is a one-input gate}
      {
        // a one-input gate\\
        \eIf{${\bf C}[g]${\rm .gate} is an M-Gate}
        {
          // apply Algorithm \ref{alg:measurement-gate}\\
          Measure($\ket{\Psi '}$,${\bf C}[g]$.$i$)\\
        }
        {
          // apply Algorithm \ref{alg:one-input-gate}\\
          QuantumOne($\ket{\Psi '}$,${\bf C}[g]$.gate,${\bf C}[g]$.$i$)\\
        }
      }
      {
        // otherwise, a two-input gate\\
        // apply Algorithm \ref{alg:two-input-gate}\\
        QuantumTwo($\ket{\Psi '}$,${\bf C}[g]$.gate,${\bf C}[g]$.$i$,${\bf C}[g]$.$j$)\\
      }
    }
    return $\ket{\Psi '}$\\
    \vspace*{1em}
    \caption{Evaluate quantum circuit}
    \label{alg:quantum-circuit}
\end{algorithm}

\section{General Architectural Decisions}
In order to accelerate the simulation process, parallelism of logic
and computational resources will be applied. In addition, rather than
creating a general computational engine, a custom architecture is
constructed specifically optimized to simulate quantum circuits. The
discussions are focused more directly on midrange and higher FPGAs.
\subsection{FPGA resources}
FPGAs provide programmable logic, ``fabric'' that can be used to
implement logic, switching, and state machines. The fabric provides
programmable units that include logic and flip-flops. In order to
support high performance data processing, digital signal processors
(DSPs) provide optimized \& flexible multipliers, adders, registers,
that provide high speed operation and efficient pipelining. A goal in
FPGA and custom design is to have fast small memories, BRAM, in the vicinity
of DSPs to reduce latencies incurred by sourcing operands. In order to
coordinate high level operation \& tasking, many FPGAs
offer hard processor cores that are typically ARM based. Other
specialized hard capabilities are available that provide high
performance in a variety of ways including data throughput (PCIe, Ethernet),
specialized AI engines, high bandwidth memories (HBMs).
\subsection{Algorithm implementation on FPGAs}
Algorithms on FPGAs can be implemented by mapping an algorithm to a
state machine that is implemented directly on the FPGA. The state
machines can be organized hierarchically in order to manage high level
states related to phases of the computation and low level state
machines that manage the state and computation on a per clock cycle
basis. 
\subsection{Permutation network}
\label{ssc:benes}
From Algorithms~\ref{alg:one-input-gate}-\ref{alg:measurement-gate},
permutation of the inputs to the functional units is
required. Notably, all algorithms perform an initial permutation to
align the QSR elements to the functional units and then an inverse
permutation to restore the initial order. Because the permutation is
common to these algorithms, the permutation operation can be factored
out, provided the permutation mechanism is general. In addition, the
observant reader can note that the restoring permutation is not
required provided the original permutation is remembered and each
subsequent permutation is performed based on the remembered
permutation. An architecture having only one permutation network
provides multiple benefits, included reducing the time necessary to
route operands to functional units and reducing resources because the
routing network can be expensive.

On a classical computer, the vector needs to be accessed
linearly and retrieving and operating on any component requires
$O(2^n)$ time. Applying a hardware solution in the form of a
permutation network can take the state vector as the input and after
passing through some number of stages in the permutation network, all
operands for a specific gate are grouped together. Non-blocking
permutation networks requiring $O(log n)$ switching layers provide an
effective solution \cite{Benes64,nassimi80}. Per the discussion in the
previous paragraph, one permutation network is required.
\subsection{One and two input gates}
The simulator will model one and two input gates that form a set of
gates supporting universal quantum computation. Gates beyond this set
are provided for convenience to simplify the composition of quantum
circuits. In order to support acceleration through parallelism, pools
of one and two gates will be available that can operate in
parallel. Furthermore, computational resources, i.e. addition and
multiplication will be restricted to one-input gates. As noted
previously, two-input gates are restricted to members of the second
order Pauli group which require neither addition nor multiplication.
\subsection{Result reporting module}
\label{ssc:rrm}
Not shown in Figure~\ref{fg:qpu-top} is the result reporting module
(RRM). Once all qubits in the circuit are measured, the QSR will have
exactly one component that is non-zero and whose magnitude is one. The
RRM will identify the index associated with that component from the
$2^n$ components in the QSR. In addition, if the QSR is not in a
suitable state, i.e. two or more states are entangled, the mechanism
will indicate it is incapable of providing a single result.
\subsection{Numerical precision}
Numerical precision can present challenges when small value
differences are important or a calculation is a result of many
intermediate steps. In an FPGA, the DSP provides optimized hard
resources to perform addition and multiplication. In addition, the
multipliers are reconfigurable so that a DSP can be reorganized into
one, two, or three multipliers of varying sizes. In this
demonstration, fixed point arithmetic is performed with an integer
precision of one bit and fractional precision of 16 bits for multiplication
operations. With this precision, a typical DSP can provide two
independent multiplications. For situations where a result is a
combination of many operands, extended precision is provided, for
example during measurement operations.
\subsection{Verification}
Verifying the operation of the quantum computing simulation
accelerator will be accomplished by comparing results with an established
quantum computing simulation platform, such as QuEST
\cite{QuEST-2019}.

\section{Quantum Simulation Unit Architecture}
\begin{figure*}[htb]
\centerline{\includegraphics[width=5in]{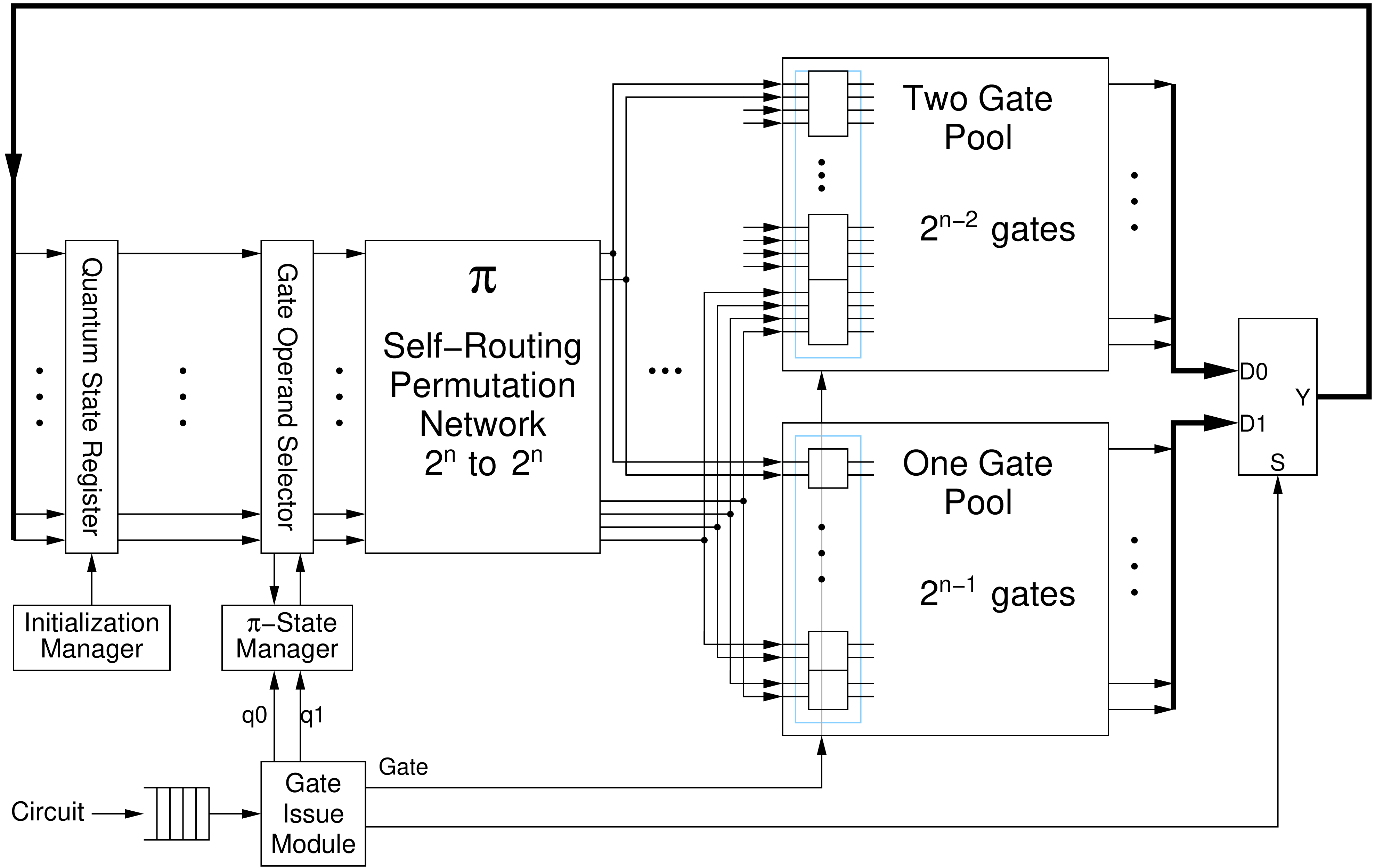}}
\caption{Top level quantum simulation unit (QSU)}
\label{fg:qpu-top}
\end{figure*}

In this section, an architecture, suitable for implementation on an
FPGA or custom silicon platform is presented. The
Figure~\ref{fg:qpu-top} gives the architecture that follows from the
discussions in the previous two sections. Omitted from this figure is
an embedded processor that is used to control \& monitor the operation
of the QSU. The demonstration has been implemented in an Intel Agilex
7 FPGA which includes a four core ARM hard processor system (HPS).

With trivial modifications, all constant one-input
gates can be easily added and all two-input gates where the magnitude
of matrix coefficients is one can be added.

An embedded processor controls the operation of the QSU implemented on
the FPGA fabric through an interface that consists of 
four 32-bit registers. The registers and their purposes are described
in Table~\ref{tb:interface}. In principal, the embedded process could
be a soft processor core or a hard processor system (HPS). In this
work, an HPS running the Yocto Linux operating system fills this role.
The \verb-HPS_control- register is a write only and controls
the overall operation of the QSU, sets the mode, configures the QSU to
receive input, and report outputs. The \verb-HPS_status- register is
read only and is used to monitor the operation of the QSU. The
\verb-HPS_data- and \verb-Fabric_data- pass data to and from the
QSU. For example, the list of simulated gates is passed from the HPS
to the fabric through \verb-HPS_data-. Simulated results and debugging
data can be sent from the QSU to the HPS through \verb-Fabric_data-.

\begin{table}[htb]
  \begin{center}
\begin{tabular}{|l|l|p{1.2in}|}
    \hline
    \multicolumn{1}{|c}{Register name} &
    \multicolumn{1}{|c}{Mode} &
    \multicolumn{1}{|c|}{Purpose} \\ \hline\hline
    \verb-HPS_control- & write only & control over fabric \\ \hline
    \verb-Fabric_status- & read only & status from fabric \\ \hline
    \verb-HPS_data- & write only & supply data to fabric \\ \hline
    \verb-Fabric_data- & read only & receive data from fabric \\ \hline
\end{tabular}
  \end{center}
  \caption{HPS QSU interface}
  \label{tb:interface}
\end{table}

The QSU architecture follows from Algorithms
Algorithms~\ref{alg:one-input-gate}-\ref{alg:quantum-circuit} with
the overarching goal to leverage their inherent parallelism. The
architecture is organized into several modules and its datapath is
shown in Figure~\ref{fg:qpu-top}. To the extent possible, modules are
designed to operate independently from the other modules and recognize
when operands and other relevant information are available to proceed
with their inherent function. The control path consists of a two level
hierarchy of state machines that controls sequencing of data through
the datapath. The top level state machine follows from
Algorithm~\ref{alg:quantum-circuit} which controls state machines for
Algorithms~\ref{alg:one-input-gate}-\ref{alg:measurement-gate}. Because
the permutation is common to both one and two inputs, that function
has been factored out and is controlled by the top level state
machine. 

\subsection{Quantum State Register (QSR)}
The QSR holds the QSV for the $n$ qubit circuit. It is an array of
$2^n$ complex fixed point values. The QSR reflects the original
ordering of qubits and not any permutation. Because many existing FPGA platforms do not
directly support floating point and to offer better utilization of
DSPs, fixed point arithmetic is 
used. 

\subsection{Initialization Manager (IM)}
\label{ssc:im}
The initialization manager provides the capability to an arbitrary
initialization of the QSR to an arbitrary entangled initial state. In
the event the initialization provided is unnormalized, the IM will
initiate normalize the state by initiating a special normalization
operation within the One-Qubit Gate Pool module.

\subsection{Gate Issue Module (GIM)}
\label{ssc:gim}
The GIM consists of two parts. The first part is a buffer and
associated buffer management that
stores the list if gates to be evaluated. The buffer is loaded through
the HPS fabric interface and new gates to be evaluated can be added at
any time. The gate issued is at the head of the buffer and is
partially decoded to retrieve the gate to be evaluated as well as its
required operands. Also included is the implementation of
Algorithm~\ref{alg:quantum-circuit}.

\subsection{Gate Operand Selector (GOS)}
\label{ssc:gos}
As required in
Algorithms~\ref{alg:one-input-gate}~and~\ref{alg:two-input-gate},
permutations are required to forward the appropriate components from the
QSR to the functional units that implement gate functions. The GOS
packages the components with current gate information, qubit inputs, and final permutation
position necessary for the permutation.

\subsection{Self-Routing Permutation Network (SRPN)}
\label{ssc:srpn}
\begin{figure}[htb]
\centerline{\includegraphics[width=\the\columnwidth]{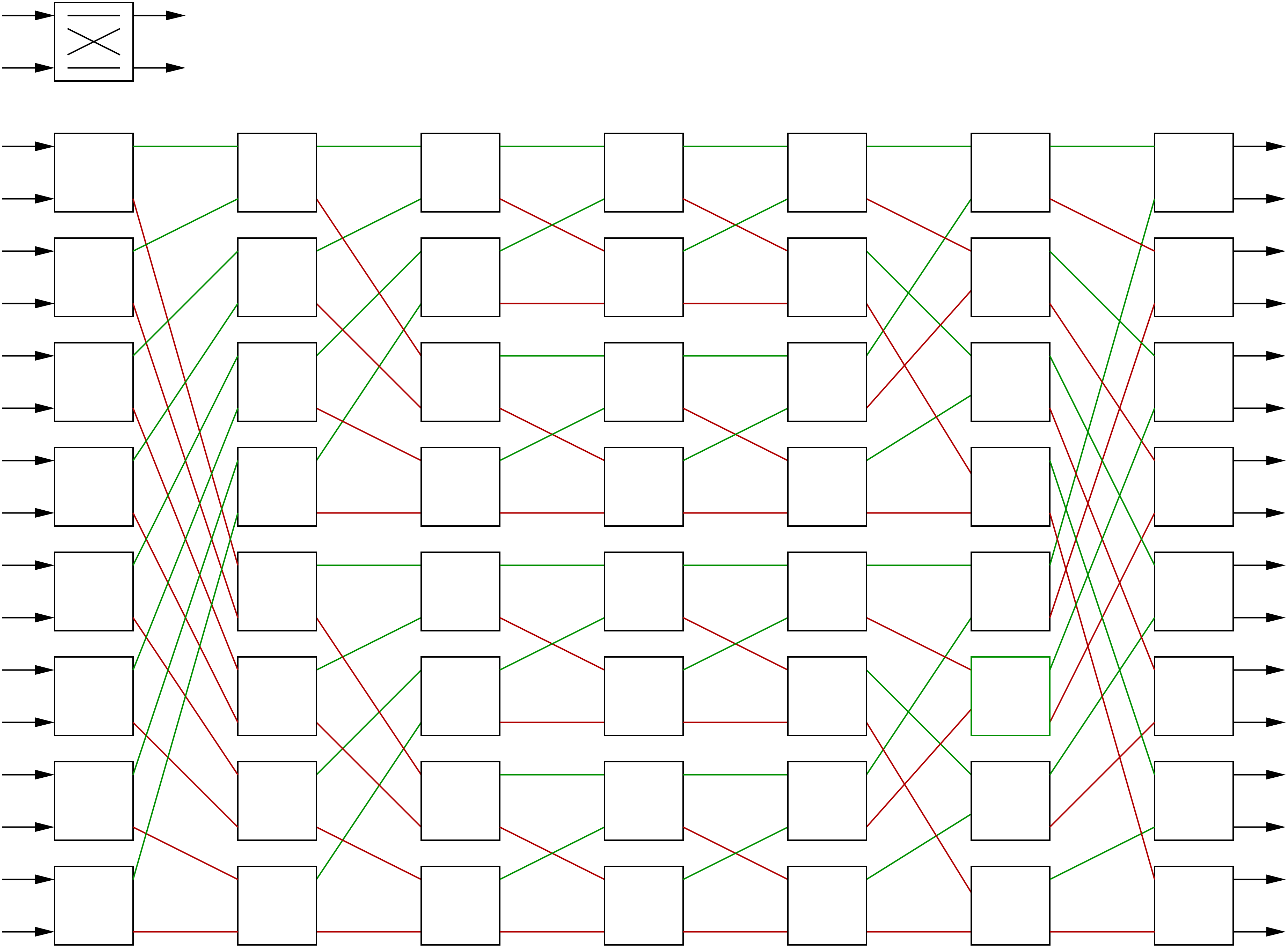}}
\caption{Self-routing permutation network for $n=4$}
\label{fg:SRPN}
\end{figure}

As noted, the permutation network must forward the operands to the
appropriate gate functional unit. Because this requires a full $2^n$
to $2^n$ permutation, an efficient permutation implementation is
required. The fastest permutation network is a crossbar switch. While
the crossbar switch can perform the permutation in one step, the
number of switches required is $2^{2n}$ which is undesirable. A
reduction in the number of switches is possible using a switching
network. While requiring more time to perform the permutation, the
number of switches is $O(2^n)$. A further challenge in performing the
permutation is ``collisions'' that result in blocking can delay
passage through the switching network. A well known non-blocking
switching network is the Bene\v{s} network \cite{Benes64} which
enables arbitrary permutations using $(n/2-1)$ layers of $2^{n-1}$
switching elements. The functionality required in the switches is
described elsewhere \cite{nassimi80}. The reader may note that the
permutation required here is not totally general and has structure
that may reduce the number of layers and will be addressed in future
work. While Algorithms \ref{alg:one-input-gate},
\ref{alg:two-input-gate}, \& \ref{alg:measurement-gate} include
permutation and inverse permutation operations, Figure~\ref{fg:qpu-top}
only shows the inclusion of one permutation network. One permutation
network is sufficient provided the permutation is known when each gate
is simulated and the original permutation is restored after the
circuit has been simulated.
An example permutation network is
shown for a $2^4$ to $2^4$ instance in Figure~\ref{fg:SRPN}. Because
arbitrary sized Bene\v{s} networks can be defined recursively,
generalized HDL specifications are possible. Furthermore, the
Bene\v{s} network is capable of partial permutations, i.e. $2^i$ to
$2^j$ where $j<i$ in the event resources are limited permitting
pipelined functional units for faster processing.

\subsection{One-Qubit Gate Pool (1-QGP)}
\label{ssc:1-qgp}
\begin{figure}[htb]
\centerline{\includegraphics[width=\the\columnwidth]{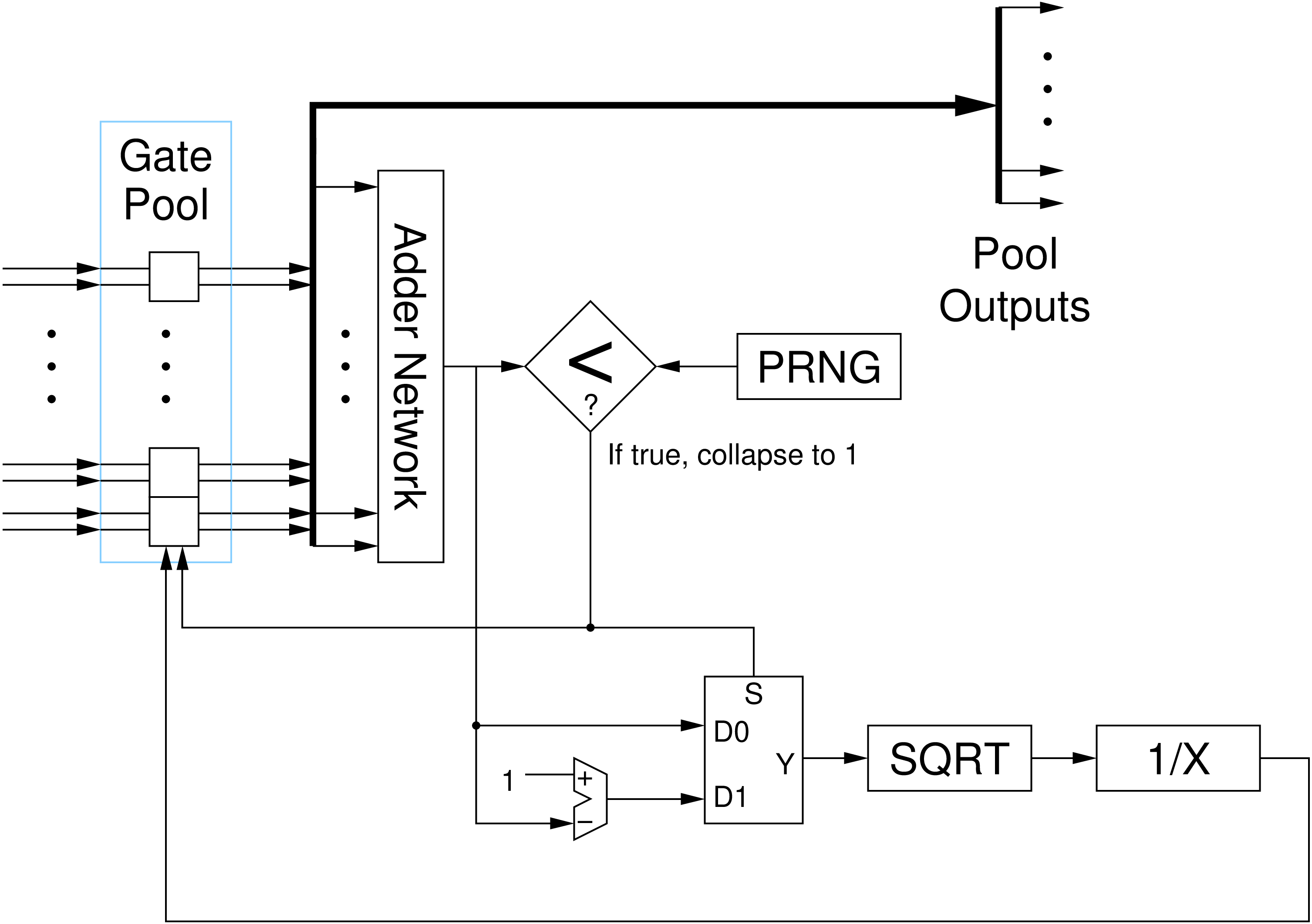}}
\caption{One qubit gate pool block diagram}
\label{fg:1-QGP}
\end{figure}

The 1-QGP holds the one-input gate functional units and controls the
operation of the gates within the pool. In order to accelerate the
simulation, gates within the pool operate in parallel on the
permuted QSR. Figure~\ref{fg:1-QGP} gives the architecture for the
gate pool.  For non-measurement gates, the results are passed on to
the network that performs the inverse permutation. For measurement
gates, applying the gate is more complex, requiring several global
operations that must be performed. First, per
Algorithm~\ref{alg:measurement-gate}, the probability that the
measurement results in 0 is calculated.
Notably, measurement gate implementation challenges were
observed by others \cite{Pilch-2019}. 
The operation requires
computing the magnitudes of the QSR components which is performed in
the one-input gate functional unites, summing the values with a
summing network within the 1-QGP. Once the probability is computed,
several special cases need to be handled. If the measured gate already
has a sharp value, no assignment of the qubit is necessary. If the
probability is neither $0$ or $1$, the qubit is collapsed by comparing
a pseudorandom number with the probability. The collapse requires that
$2^{n-1}$ components be zeroed as a result of one qubit value now
being excluded from possibility. At this point, the QSR is no longer
properly normalized, and the square root of the winning bit's
probability must be computed followed by computing its inverse.

Because of the likelihood of numerical errors, computed probabilities
may not sum exactly to precise values as implied by the
mathematics. This situation is mitigated, but not entirely eliminated,
by using extended precision arithmetic in the adder network, doubling
the precision of the fixed point number and including an error
tolerance that is a function of the number of qubits in the circuit.
For the inverse calculation, because the inverse is greater than 1, an
extended precision type capable of holding these values is used.

As noted for universal computation, in support of the minimum set of
quantum gates required, the one-input Hadamard and T~gates are
required. Further, the measurement or M~gate is required to retrieve
answers. For convenience to support a wide variety of quantum
circuits, many more gates have been implemented and are listed in
Table~\ref{tb:oneInputGates} where $j=\sqrt{-1}$.

The required arithmetic operations local to each gate are addition and
multiplication. Rotation gates are common in many quantum
algorithms. Any constant rotation can be included by specifying its
unitary matrix. At the present, the simulation is not capable of
handling an arbitrary matrix. If this is desired, a simple fix would
be to augment the HPS-fabric interface to permit the HPS to provide
any desired unitary matrix.

\newcommand{\VinvRootTwo}{$\frac{1}{\sqrt{2}} $ }
\newcommand{\VinvTwo}{$\frac{1}{2} $ }
\newcommand{\VpiByTwo}{$\frac{\pi}{2} $ }
\newcommand{\VpiByFour}{$\frac{\pi}{4} $ }
\newcommand{\VrotPpiByTwo}{$e^{j\VpiByTwo}$ }
\newcommand{\VrotPpiByFour}{$e^{j\VpiByFour}$ }

\newcommand{\RootXtl}{(1+j)}
\newcommand{\RootXtr}{(1-j)}
\newcommand{\RootXbl}{\RootXtr}
\newcommand{\RootXbr}{\RootXtl}

\newcommand{\RootYtl}{(1+j)}
\newcommand{\RootYtr}{(-1-j)}
\newcommand{\RootYbl}{\RootYtl}
\newcommand{\RootYbr}{\RootYtl}

\newcommand{\MatStrut}{\rule[-\baselineskip]{0pt}{2.4\baselineskip}}

\newcommand{\PauliX}{\MatStrut%
$\left[ \begin{array}{ll}0 & 1 \\ 1 & 0 \end{array} \right] $ }
\newcommand{\PauliY}{\MatStrut$ \left[ \begin{array}{rr}0 & j \\ -j & 0 \end{array} \right] $ }
\newcommand{\PauliZ}{\MatStrut$ \left[ \begin{array}{rr}1 & 0 \\ 0 & -1 \end{array} \right] $ }
\newcommand{\Hgate}{\MatStrut%
  \VinvRootTwo $
  \left[
    \begin{array}{rr}
      1 & 1 \\
      1 & -1
    \end{array}
    \right]
  $
}
\newcommand{\Sgate}{\MatStrut$ \left[ \begin{array}{rr}1 & 0 \\ 0 & j \end{array} \right] $ }

\newcommand{\Tgate}{\MatStrut$ \left[ \begin{array}{rl}1 & 0 \\ 0 & e^{j\pi /4}\end{array} \right] $ }

\newcommand{\Vgate}{\MatStrut%
  \VinvTwo $ \left[
    \begin{array}{ll}
      \RootXtl & \RootXtr \\
      \RootXbl & \RootXbr
    \end{array} \right] $%
}
\newcommand{\NopGate}{\MatStrut$ \left[ \begin{array}{rr}1 & 0 \\ 0 & 1\end{array} \right] $ }
\newcommand{\RootYGate}{\MatStrut%
  \VinvTwo $ \left[
    \begin{array}{lr}
      \RootYtl & \RootYtr \\
      \RootYbl & \RootYbr
    \end{array} \right] $%
}

\newcommand{\SInvGate}{\MatStrut$ \left[ \begin{array}{rr}1 & 0 \\ 0 & -j \end{array} \right] $ }
\newcommand{\TInvGate}{\MatStrut$ \left[ \begin{array}{rl}1 & 0 \\ 0 & e^{-j\pi
        /4}\end{array} \right] $ }

\newcommand{\ErrGate}{$ \left[ \begin{array}{rr}0 & j \\ -j & 0 \end{array} \right] $ }
\newcommand{\ErrXGate}{$ \left[ \begin{array}{rr}0 & j \\ -j & 0 \end{array} \right] $ }
\newcommand{\ErrYGate}{$ \left[ \begin{array}{rr}0 & j \\ -j & 0 \end{array} \right] $ }
\newcommand{\ErrZGate}{$ \left[ \begin{array}{rr}0 & j \\ -j & 0 \end{array} \right] $ }

\begin{table}[htb]
  \begin{center}
    \begin{footnotesize}
\begin{tabular}{|c|c|p{1.75in}|}
\hline
\bf Gate   & \bf Name                        & \bf Function \\ \hline \hline
Nop        & No-op gate                      & \NopGate     \\ \hline
$X$        & Pauli X                         & \PauliX      \\ \hline
$Y$        & Pauli Y                         & \PauliY      \\ \hline
$Z$        & Pauli Z                         & \PauliZ      \\ \hline
$H$        & Hadamard                        & \Hgate       \\ \hline
$V$        & $\sqrt{\mbox{NOT}}$, $\sqrt{X}$ & \Vgate       \\ \hline
$\sqrt{Y}$ &                                 & \RootYGate   \\ \hline
$S$        &  $\sqrt{Z}$                     & \Sgate       \\ \hline
$S^{-1}$    &  Inverse of $S$                 & \SInvGate   \\ \hline
$T$        &  $\sqrt{S}$                     & \Tgate       \\ \hline
$T^{-1}$    &  Inverse of $T$                 & \TInvGate    \\ \hline
$E_x$      & Error in X & Apply Pauli X with a designated error probability
\\ \hline
$E_y$ & Error in X & Apply Pauli Y with a designated error probability
\\ \hline
$E_z$ & Error in Z & Apply Pauli Z with a designated error probability
\\ \hline
$M$ & Measurement gate & Measures and collapses the state for the
specified qubit
\\ \hline
\end{tabular}
\end{footnotesize}
\end{center}
\caption{Implemented one-input quantum gates}
\label{tb:oneInputGates}
\end{table}

\subsection{Two-Qubit Gate Pool (2-QGP)}
\label{ssc:2-qgp}
The 2-QGP models two-input gates in a similar fashion to one-input
gates. The permutation network routing differs because the two-input
gates require four components from the QSR. As noted,
multiplications are restricted to the one-input gates, limiting the
nature of two-input gates. The unitary matrices defining two-input gates
requires matrix elements be restricted to $\{1,-1,j,-j\}$.
A summary of the two-input gates implemented are shown in
Table~\ref{tb:twoInputGates}.  Overall, the architecture of 2-QGP is
relatively simple compared with the 1-QGP.
\newcommand{\MatTwoStrut}{\rule[-1.4\baselineskip]{0pt}{3.25\baselineskip}}
\newcommand{\CPauliY}{ \MatTwoStrut \small $\scriptsize \left[
    \begin{array}{rrrr}
      1 & 0 & 0 & 0 \\
      0 & 1 & 0 & 0 \\
      0 & 0 & 0 & j \\
      0 & 0 & -j & 0
    \end{array}
    \right]
  $%
}
\newcommand{\CPauliX}{%
  \MatTwoStrut%
  $\scriptsize
  \left[
    \begin{array}{rrrr}
      1 & 0 & 0 & 0 \\
      0 & 1 & 0 & 0 \\
      0 & 0 & 0 & 1 \\
      0 & 0 & 1 & 0
    \end{array}
    \right]
  $%
}
\newcommand{\CPauliZ}{%
  \MatTwoStrut%
  $\scriptsize
  \left[
    \begin{array}{rrrr}
      1 & 0 & 0 & 0 \\
      0 & 1 & 0 & 0 \\
      0 & 0 & 1 & 0 \\
      0 & 0 & 0 & -1
    \end{array}
    \right]
  $%
}
\newcommand{\SqrtZZ}{%
  \MatTwoStrut%
  $\scriptsize
  \left[
    \begin{array}{rrrr}
      1 & 0 & 0 & 0 \\
      0 & j & 0 & 0 \\
      0 & 0 & j & 0 \\
      0 & 0 & 0 & -1
    \end{array}
    \right]
  $%
}
\newcommand{\Swap}{%
  \MatTwoStrut%
  $\scriptsize
  \left[
    \begin{array}{rrrr}
      1 & 0 & 0 & 0 \\
      0 & 0 & 1 & 0 \\
      0 & 1 & 0 & 0 \\
      0 & 0 & 0 & 1
    \end{array}
    \right]
  $%
}

\begin{table}[htb]
  \begin{center}
\begin{tabular}{|c|c|p{1in}|}
\hline
\bf Gate & \bf Name & \bf Function  \\ \hline \hline
$\mathit{CNOT}$     & Controlled-NOT     & \CPauliX \\ \hline
$\mathit{CY}$       & Controlled Pauli Y & \CPauliY \\ \hline
$\mathit{CZ}$       & Controlled Pauli Z & \CPauliZ \\ \hline
\rule{0pt}{10pt}$\sqrt{\mathit{ZZ}}$& $\sqrt{Z\bigotimes Z}$   & \SqrtZZ \\ \hline
$\mathit{SWAP}$     & Swap gate          & \Swap \\ \hline
\end{tabular}

\end{center}
  \caption{Two-input gates}
  \label{tb:twoInputGates}
\end{table}

\begin{figure}[htb]
\centerline{\includegraphics[scale=0.25]{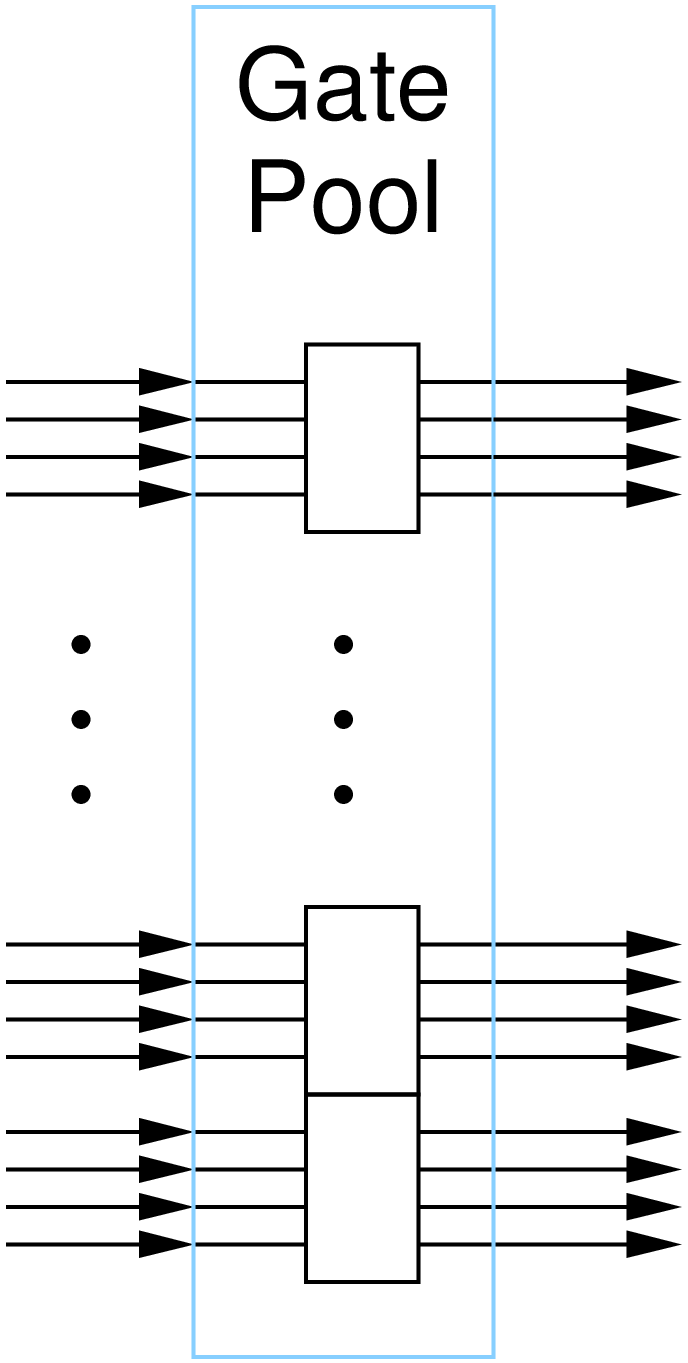}}
\caption{Two qubit gate pool block diagram}
\label{fg:2-QGP}
\end{figure}

\subsection{Universal Set of Gates}
A review of Tables~\ref{tb:oneInputGates}~and~\ref{tb:twoInputGates} reveals that
the QSU includes gates that support a universal set \cite{Boykin99}
since CNOT~gate, Hadamard Gate and the T~are included in the set of
gates implemented.

\newcommand{\Reg}{\textsuperscript{\sffamily\textregistered}}

\section{Example}
The QSU is implemented at the register transfer language (RTL) level
using 2008 standard release of the VHSIC Hardware Description Language
(VHDL) \cite{VHDL-LRM-2008}.  The QSU is verified through simulation
and also on an FPGA for two circuits and the results are described in
this section.

\subsection{Simulation Results}
Simulation was performed first on the RTL model and then confirmed
using the QuEST \cite{QuEST-2019} quantum computing simulator. The
results for two circuits are given here.

\subsubsection{Simple Circuit.}
\begin{figure}[htb]
\centerline{\includegraphics[width=\the\columnwidth]{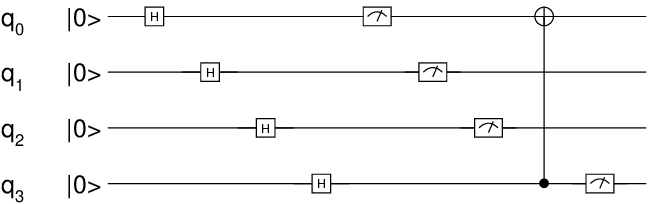}}
\caption{Test circuit schematic}
\label{fg:test-circuit}
\end{figure}

VHDL simulation results have been performed
using the open source VHDL platform GHDL v2.0 \cite{GHDL-2} for a
simple circuit composed of
a layer of Hadamard gates to fully entangle the quantum state
is applied. This is followed by a measurement of one qubit, a
controlled-NOT gate, and then the subsequent measurement of the
remaining qubits. The circuit is shown in Figure~\ref{fg:test-circuit}.
\subsubsection{Random Circuit.}
\begin{figure*}[htb]
  \centerline{\includegraphics[width=6.5in]{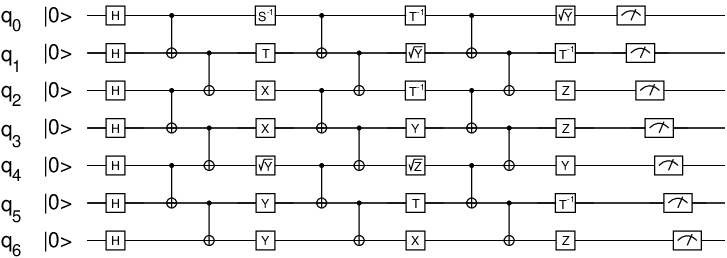}}
    \caption{Random circuit}
\label{fg:randCircuit}
\end{figure*}

Random circuits provide a stronger verification of the QSU
architecture's correct operation because biases are eliminated in
the creation of the circuit.  In addition, random quantum circuits
have benefit in their own right in regards to solving certain classes of
problems (see for example \cite{Emerson04,Fisher23}.  A 7-qubit random
circuit example is shown in Figure~\ref{fg:randCircuit} using an
approach outlined in \cite{Emerson04}.  The circuit is generated by
first applying the Hadamard gate to each input, fully entangling all
inputs.  Next, for three iterations, $\mathit{CNOT}$ gates are applied
as shown followed by, for each qubit, gates randomly selected from the
list $\{X,Y,Z,S,S^{-1},T,T^{-1}, \sqrt{X}, \sqrt{Y}, \sqrt{Z}\}$. In
Figure~\ref{fg:randCircuit}, the staggered order of the gates reflects
their simulation order.  Once the circuit has been defined, it is
included in a test bench for simulation of the QSU HDL specification.
Aggregate statistics are generated regarding the final output of
the circuit for ten thousand trials. Next, the simulation is performed
using QuEST. More simulation trials, one million, are run to achieve a
more refined estimate of the expected distribution among final
states. Finally, the QSU is programmed onto the FPGA and the circuit
is similarly run for one million trials. The results are summarized in
Table~\ref{tb:verification-example}.

Qualitatively comparing the QuEST and QSU FPGA based results, the
respective empirically derived state probabilities are very similar
and are zero for the same quantum states. Further, the Euclidean
distance between the distributions is $1.45\times 10^{-4}$, mean
absolute error
between the distributions is $1.36\times 10^{-5}$ with a standard
deviation of $1.21\times 10^{-5}$. These suggest very good agreement
between the two simulation results.

\newcommand{\PowStrut}{\rule[-2\baselineskip]{1pt}{5\baselineskip}}

\begin{table}[htb]
  \begin{center}
    \begin{footnotesize}
  \begin{tabular}{|l@{\hspace{0.25em}}|rl@{\hspace{0.25em}}|rl@{\hspace{0.25em}}|rl@{\hspace{0.25em}}|}
    \hline
    \multicolumn{1}{|c|}{\bf State} 
    & \multicolumn{2}{c|}{\bf QSU (HDL)} &
    \multicolumn{2}{c|}{\bf QuEST} &
    \multicolumn{2}{c|}{\bf QSU (FPGA)} \\ \cline{2-7} \cline{2-7}
    \multicolumn{1}{|c|}{$\ket{\psi}$} 
    & \multicolumn{1}{c}{\rule{0pt}{10pt} n/$10^4$} & 
    \multicolumn{1}{c|}{$\mu_{\ket{\psi}}$} & 
    \multicolumn{1}{c}{n/$10^6$} & 
    \multicolumn{1}{c|}{$\mu_{\ket{\psi}}$} & 
    \multicolumn{1}{c}{n/$10^6$} & 
    \multicolumn{1}{c|}{$\mu_{\ket{\psi}}$} \\ \hline
    $\ket{\mbox{\tt 0x10}}$ &  281 &  0.00281  &  23738 & 0.00237  &  23420 & 0.00234  \\ 
    $\ket{\mbox{\tt 0x11}}$ &   76 &  0.00076  &   7869 & 0.00079  &   7938 & 0.00079  \\ 
    $\ket{\mbox{\tt 0x12}}$ &   84 &  0.00084  &   7792 & 0.00078  &   7895 & 0.00079  \\ 
    $\ket{\mbox{\tt 0x13}}$ &  252 &  0.00252  &  23489 & 0.00234  &  23697 & 0.00237  \\ 
    $\ket{\mbox{\tt 0x14}}$ &   86 &  0.00086  &   7797 & 0.00078  &   7888 & 0.00079  \\ 
    $\ket{\mbox{\tt 0x15}}$ &  233 &  0.00233  &  23478 & 0.00234  &  23587 & 0.00236  \\ 
    $\ket{\mbox{\tt 0x16}}$ &  217 &  0.00217  &  23287 & 0.00232  &  23493 & 0.00235  \\ 
    $\ket{\mbox{\tt 0x17}}$ &   74 &  0.00074  &   7912 & 0.00079  &   7971 & 0.00080  \\ 
    $\ket{\mbox{\tt 0x18}}$ &  229 &  0.00229  &  23420 & 0.00234  &  23412 & 0.00234  \\ 
    $\ket{\mbox{\tt 0x19}}$ &   76 &  0.00076  &   7814 & 0.00078  &   7830 & 0.00078  \\ 
    $\ket{\mbox{\tt 0x1a}}$ &   77 &  0.00077  &   7783 & 0.00078  &   7739 & 0.00077  \\ 
    $\ket{\mbox{\tt 0x1b}}$ &  225 &  0.00225  &  23348 & 0.00233  &  23384 & 0.00234  \\ 
    $\ket{\mbox{\tt 0x1c}}$ &   85 &  0.00085  &   7690 & 0.00077  &   7880 & 0.00079  \\ 
    $\ket{\mbox{\tt 0x1d}}$ &  249 &  0.00249  &  23348 & 0.00233  &  23676 & 0.00237  \\ 
    $\ket{\mbox{\tt 0x1e}}$ &  224 &  0.00224  &  23703 & 0.00237  &  23803 & 0.00238  \\ 
    $\ket{\mbox{\tt 0x1f}}$ &   82 &  0.00082  &   7795 & 0.00078  &   7836 & 0.00078  \\ 
    $\ket{\mbox{\tt 0x30}}$ &  229 &  0.00229  &  23495 & 0.00235  &  23627 & 0.00236  \\ 
    $\ket{\mbox{\tt 0x31}}$ &   87 &  0.00087  &   7755 & 0.00076  &   7677 & 0.00077  \\ 
    $\ket{\mbox{\tt 0x32}}$ &   92 &  0.00092  &   7840 & 0.00078  &   7835 & 0.00078  \\ 
    $\ket{\mbox{\tt 0x33}}$ &  240 &  0.00240  &  23369 & 0.00234  &  23470 & 0.00235  \\ 
    $\ket{\mbox{\tt 0x34}}$ &   74 &  0.00074  &   7762 & 0.00078  &   7735 & 0.00077  \\ 
    $\ket{\mbox{\tt 0x35}}$ &  224 &  0.00224  &  23227 & 0.00232  &  23466 & 0.00235  \\ 
    $\ket{\mbox{\tt 0x36}}$ &  231 &  0.00231  &  23449 & 0.00234  &  23364 & 0.00234  \\ 
    $\ket{\mbox{\tt 0x37}}$ &  100 &  0.00100  &   7790 & 0.00079  &   7836 & 0.00078  \\ 
    $\ket{\mbox{\tt 0x38}}$ &  215 &  0.00215  &  23378 & 0.00234  &  23563 & 0.00236  \\ 
    $\ket{\mbox{\tt 0x39}}$ &   74 &  0.00074  &   7840 & 0.00078  &   7773 & 0.00078  \\ 
    $\ket{\mbox{\tt 0x3a}}$ &   76 &  0.00076  &   7766 & 0.00078  &   7790 & 0.00078  \\ 
    $\ket{\mbox{\tt 0x3b}}$ &  253 &  0.00253  &  23312 & 0.00233  &  23374 & 0.00234  \\ 
    $\ket{\mbox{\tt 0x3c}}$ &   74 &  0.00074  &   7678 & 0.00077  &   7761 & 0.00078  \\ 
    $\ket{\mbox{\tt 0x3d}}$ &  224 &  0.00224  &  23451 & 0.00235  &  23429 & 0.00234  \\ 
    $\ket{\mbox{\tt 0x3e}}$ &  228 &  0.00228  &  23560 & 0.00236  &  23292 & 0.00233  \\ 
    $\ket{\mbox{\tt 0x3f}}$ &   73 &  0.00073  &   7727 & 0.00077  &   7989 & 0.00080  \\ 
    $\ket{\mbox{\tt 0x40}}$ &  224 &  0.00224  &  23380 & 0.00234  &  23484 & 0.00235  \\ 
    $\ket{\mbox{\tt 0x41}}$ &   81 &  0.00081  &   7840 & 0.00078  &   7878 & 0.00079  \\ 
    $\ket{\mbox{\tt 0x42}}$ &   76 &  0.00076  &   7837 & 0.00078  &   7868 & 0.00079  \\ 
    $\ket{\mbox{\tt 0x43}}$ &  238 &  0.00238  &  23427 & 0.00234  &  23376 & 0.00234  \\ 
    $\ket{\mbox{\tt 0x44}}$ &   82 &  0.00082  &   7829 & 0.00078  &   7757 & 0.00078  \\ 
    $\ket{\mbox{\tt 0x45}}$ &  228 &  0.00228  &  23513 & 0.00235  &  23643 & 0.00236  \\ 
    $\ket{\mbox{\tt 0x46}}$ &  227 &  0.00227  &  23325 & 0.00233  &  23519 & 0.00235  \\ 
    $\ket{\mbox{\tt 0x47}}$ &   83 &  0.00083  &   7801 & 0.00078  &   7797 & 0.00078  \\ 
    $\ket{\mbox{\tt 0x48}}$ &  247 &  0.00247  &  23621 & 0.00236  &  23460 & 0.00235  \\ 
    $\ket{\mbox{\tt 0x49}}$ &   75 &  0.00075  &   7867 & 0.00079  &   7678 & 0.00077  \\ 
    $\ket{\mbox{\tt 0x4a}}$ &   86 &  0.00086  &   7805 & 0.00078  &   7723 & 0.00077  \\ 
    $\ket{\mbox{\tt 0x4b}}$ &  232 &  0.00232  &  23349 & 0.00233  &  23295 & 0.00233  \\ 
    $\ket{\mbox{\tt 0x4c}}$ &   54 &  0.00054  &   7802 & 0.00078  &   7780 & 0.00078  \\ 
    $\ket{\mbox{\tt 0x4d}}$ &  215 &  0.00215  &  23661 & 0.00237  &  23438 & 0.00234  \\ 
    $\ket{\mbox{\tt 0x4e}}$ &  248 &  0.00248  &  23174 & 0.00232  &  23330 & 0.00233  \\ 
    $\ket{\mbox{\tt 0x4f}}$ &   68 &  0.00068  &   7835 & 0.00078  &   7704 & 0.00077  \\ 
    $\ket{\mbox{\tt 0x60}}$ &  246 &  0.00246  &  23137 & 0.00231  &  23451 & 0.00235  \\ 
    $\ket{\mbox{\tt 0x61}}$ &   75 &  0.00075  &   7736 & 0.00077  &   7885 & 0.00079  \\ 
    $\ket{\mbox{\tt 0x62}}$ &   96 &  0.00096  &   7851 & 0.00079  &   7748 & 0.00077  \\ 
    $\ket{\mbox{\tt 0x63}}$ &  242 &  0.00242  &  23271 & 0.00233  &  23320 & 0.00233  \\ 
    $\ket{\mbox{\tt 0x64}}$ &   84 &  0.00084  &   7874 & 0.00079  &   7718 & 0.00077  \\ 
    $\ket{\mbox{\tt 0x65}}$ &  265 &  0.00265  &  23448 & 0.00234  &  23398 & 0.00234  \\ 
    $\ket{\mbox{\tt 0x66}}$ &  225 &  0.00225  &  23447 & 0.00234  &  23228 & 0.00232  \\ 
    $\ket{\mbox{\tt 0x67}}$ &   83 &  0.00083  &   7960 & 0.00080  &   7783 & 0.00078  \\ 
    $\ket{\mbox{\tt 0x68}}$ &  217 &  0.00217  &  23560 & 0.00236  &  23230 & 0.00232  \\ 
    $\ket{\mbox{\tt 0x69}}$ &   59 &  0.00059  &   7828 & 0.00078  &   7750 & 0.00078  \\ 
    $\ket{\mbox{\tt 0x6a}}$ &   78 &  0.00078  &   7761 & 0.00078  &   7791 & 0.00078  \\ 
    $\ket{\mbox{\tt 0x6b}}$ &  237 &  0.00237  &  23345 & 0.00233  &  23489 & 0.00235  \\ 
    $\ket{\mbox{\tt 0x6c}}$ &   67 &  0.00067  &   8057 & 0.00081  &   7741 & 0.00077  \\ 
    $\ket{\mbox{\tt 0x6d}}$ &  213 &  0.00213  &  23524 & 0.00235  &  23434 & 0.00234  \\ 
    $\ket{\mbox{\tt 0x6e}}$ &  234 &  0.00234  &  23546 & 0.00235  &  23098 & 0.00231  \\ 
    $\ket{\mbox{\tt 0x6f}}$ &   71 &  0.00071  &   7927 & 0.00079  &   7776 & 0.00078  \\ 
    others                  &    0 &  0.00000  &      0 & 0.00000  &      0 & 0.00000  \\ \hline
   
  \end{tabular}
  \end{footnotesize}
  \end{center}
  \caption{Summary of verification results, rounded to five places}
  \label{tb:verification-example}
\end{table}

\subsection{FPGA Demonstration}
The QSU has been implemented and evaluated on an Intel\Reg Agilex\Reg
AGFB014R24A2E2VR0 (AGFB014) FPGA on the Intel\Reg Agilex\Reg F-Series
FPGA development kit \cite{Intel-Agilex}. The FPGA programming has
been developed using the Intel\Reg Quartus\Reg Prime Pro version 21.1
development platform \cite{Intel-Quartus}. The AGFB014 fabric includes
487,200 Adaptive Logic Modules (ALMs) (1,437,240 logic element
equivalent), 145.6Mb of BRAM, 4,510 digital signal processors
(DSPs). In addition, the AGFB014 includes hard IP Ethernet capable of
supporting data rates up to 400Gbps data rates and PCIe 4.0.  In
addition, the HPS is a quad core 64-bit Arm Cortex-A53.

Table~\ref{tb:fpga-stats} provides a summary of the utilization of
some major QSU modules for a QSU capable of simulating a 7-qubit circuit.
\begin{table}[htb]
  \begin{center}
    \begin{tabular}{|l|r@{\hspace{5pt}}l|}
      \hline
      \multicolumn{1}{|c}{\bf Model} &
      \multicolumn{2}{|c|}{\bf Utilization}
      \\ \hline \hline
      \multirow{2}{*}{1-QGP~~~~~~~~}
      & 72,894 & ALMs \\ \cline{2-3}
      & 1,600 & DSPs \\ \hline
      2-QGP      & 8,333  & ALMs \\ \hline
      $\pi$     & 61,298 & ALMs \\ \hline
      GIM/GOS          & 505     & ALMs \\ \hline
      RRM       & 1,328   & ALMs \\ \hline
      miscellaneous       & 21,232   & ALMs \\ \hline\hline
      Total       & 168,931 & ALMs \\ \hline
    \end{tabular}
  \end{center}
  \caption{Module utilization}
\label{tb:fpga-stats}
\end{table}

\subsection{Performance}
Few attempts were made to optimize the performance and space of the
QSU and, rather, creating a working prototype was prioritized.  Making
it is challenging to make direct comparisons.  Simulating the random
circuit required 1,430 clocks per individual circuit simulation. With
a clock period of 10ns, the time required for an individual simulation
is 14.3$\mu$s Notably, simulating measurement gates consumed 271
clocks, or nearly 20\% of the simulation clock cycles.  For
comparison, the random circuit simulation using QuEST on a XEON
processor (XEON ES-2603 v2, clock frequency of 1.8GHz) requires 5.25ms
of CPU time/simulation. The FPGA based simulation provides a speedup
of 368.

\section{Summary and Future Work}
In this paper, an FPGA based quantum computing simulator was
presented. The simulator models sufficient gates to support universal
quantum computation. Further, the simulator architecture was conceived
to be extensible and scalable. The simulator was modeled at the RTL
level using VHDL. A demonstration implementation on an Intel Agilex
FPGA was capable of simulating a seven qubit quantum circuit. The FPGA
results were verified by comparing with simulations both of the VHDL
implementation and with respect to the QuEST quantum computing
simulator.

Several potential directions for future work are considered. First,
the model presented here has been implemented as a proof of concept
and beyond architectural features \& use of massive parallelism, few
attempts were made to optimize performance \& resources. Of course,
simulating an arbitrary quantum circuit is at least linear in time \&
resources with respect to the length of the state vector. The time
required to perform certain serial aspects, such as the square root
\& division required in the measurement gate can be improved. In
addition, the number of DSPs required for each individual quantum gate
can potentially increase the potential parallelism. Second,
demonstrating scalability across multiple devices would enable
simulating larger problem sizes. Third, the quantum circuit simulation
can be coupled with classical computing algorithms to simulate
algorithms that include both classical procedural and quantum
aspects. Fourth, additional flexibility can be provided by providing a
capability of loading arbitrary unitary matrices for one-input
gates. For example, algorithms such as the quantum Fourier
transform (QFT) \cite{coppersmith02} require several different phase
gates, as determined by the size of the QFT, which are not directly
supported in the QSU. 

\bibliographystyle{SageH}
\def\UrlBreaks{\do\/\do-}

\bibliography{quantum}

\end{document}